\documentclass[twocolumn,showpacs,aps,prl]{revtex4}
\usepackage{graphicx,amsmath,amssymb}
\usepackage{dcolumn,fancyhdr}
\usepackage{bm,natbib}
\usepackage{times}
\usepackage{txfonts}
\usepackage{epstopdf}
\usepackage{latexsym}
\usepackage{epsfig}

\newcommand{\ket}[1]{\left\vert#1\right\rangle}

\newcommand{\bra}[1]{\left\langle#1\right\vert}

\newcommand{\theo}[1]{{\bf Theorem.\,} {\it #1}}

\usepackage{color}
\definecolor{Blue}{rgb}{0,0,1}
\definecolor{Red}{rgb}{1,0,0}
\definecolor{Green}{rgb}{0,1,0}
\definecolor{Purp}{rgb}{.2,0,.2}
\definecolor{white}{rgb}{1,1,1}

\begin{document}
\title{Dynamical role of system-environment correlations in non-Markovian dynamics}
\author{Laura Mazzola$^1$, C\'esar A. Rodr\'iguez-Rosario$^{2,3}$, Kavan Modi$^4$, and Mauro Paternostro$^1$}
\affiliation{$^1$Centre for Theoretical Atomic, Molecular and Optical Physics, School of Mathematics and Physics, Queen's University Belfast, BT7 1NN Belfast, United Kingdom\\
$^2$Bremen Center for Computational Materials Science, Universit\"at Bremen, Am Fallturm 1, 28359 Bremen, Germany\\
$^3$Department of Chemistry and Chemical Biology, Harvard University, Cambridge MA, USA\\
$^4$Centre for Quantum Technologies, National University of Singapore, Singapore}

\begin{abstract}
We analyse the role played by system-environment correlations in the emergence of non-Markovian dynamics. By working within the framework developed in Breuer {\it et al.}, Phys. Rev. Lett. \textbf{103}, 210401 (2009), we unveil a fundamental connection between non-Markovian behaviour and dynamics of system-environment correlations. We derive an upper bound to the rate of change of the distinguishability between different states of the system 
that explicitly depends on the establishment of correlations between system and environment. We illustrate our results using a fully solvable spin-chain model, which allows us to gain insight on the mechanisms triggering non-Markovian evolution. 
\end{abstract}

\maketitle

The study of non-Markovian quantum dynamics is gathering substantial interest due to key advances in the analysis, understanding and even simulation of non-trivial system-environment effects~\cite{guo}. A particularly significant step forward in this context has been performed with the formulation of new theoretical tools able to characterize and quantify the {\it deviations} of given dynamics from Markovianity~\cite{WolfPRL08,RivasPRL10,BreuerPRL09,Fisher,RajagopalPRA10}. Each of such instruments addresses a specific manifestation of non-Markovianity, therefore embodying {\it in principle} a distinct quantitative measure. Although the relationship among such diversified approaches has been considered~\cite{HaikkaPRA11} and a full reconciliation seems foreseeable~\cite{ChruscinskiPRA11}, the fundamental responsibles for the occurrence of non-Markovian features remain largely elusive.

In particular the question of whether system-environment correlations (SECs) are of any importance in the emergence of non-Markovian dynamics is essentially {\it open}. Although a possible role of SECs was hinted in some formulation of non-Markovian dynamical maps~\cite{NonMarko}, their connection with quantitative measures of non-Markovianity has not been explored, to the best of our knowledge. Here, we aim at filling this gap by formulating a theory that makes such a connection explicit and indeed experimentally testable. We show that, by making use of the properties of the trace distance, it is possible to establish a quantitative link between the non-Markovian nature of a process and the existence/evolution of dynamical SECs. 
Our key result is the formulation of an upper bound for the derivative of the trace distance between two evolving states of the system that depends explicitly on SECs and environmental distinguishability. In turn, this bound can be used to witness the occurrence of SECs by monitoring experimentally the behavior of the trace distance. The development of increasingly accurate and reliable techniques for manipulation of quantum systems in photonics and condensed-matter physics will soon open up the possibility to harness the interaction between system and environment. In this perspective, our results provide an analytic tool to gather useful information on the way a system and its environment share quantum correlations by monitoring a figure of merit of simple experimental access.

In order to link our formal findings to an interesting physical case, we use an exactly solvable quantum many-body model. We consider a quantum spin chain ruled by XX-like inter-particle couplings in a transverse magnetic field, a system that has been recently used to ascertain non-Markovian dynamical features~\cite{ApollaroPRA11}. This provides a pragmatic scenario where we illustrate our results: on one hand, we test the tightness of the bound using a physically motivated example. On the other hand, we quantify the amount of SECs and show the similarity between the behavior of the rate of change of SECs and that of the derivative of the trace distance. As we discuss thoroughly, our investigation almost naturally paves the way to the critical assessment of the definition of ``information flow" upon which the measure in~\cite{BreuerPRL09} is built.

\emph{Trace distance-based measure of non-Markovianity.}---We start by setting up a framework for relating non-Markovianity and SECs and based on state distinguishability, which is measured,  throughout this paper, by the trace distance $D(\rho_1,\rho_2)=||\rho_1-\rho_2||/2$ between two states $\rho_1$ and $\rho_2$. Here, $||A||=\mathrm{Tr}\sqrt{A^{\dag}A}$ is the trace-$1$ norm of a matrix $A$. 
Given a quantum system that is prepared in state $\rho_j~(j=1,2)$ with probability $1/2$, the average probability for an observer to guess correctly such preparation is $[1+D(\rho_1,\rho_2)]/2$. 
In what follows, we exploit the subadditive property of the trace norm with respect to the tensor product, {\it i.e.} $D(\rho_1{\otimes}\rho_2,\rho_3{\otimes}\rho_4){\leq}D(\rho_1,\rho_3){+}D(\rho_2,\rho_4)$ with $\rho_{k}$ ($k{=}1,..,4$) arbitrary density matrices, and its contractivity under a positive trace-preserving map $\Psi_t$
$D(\Psi_{t}[\rho_1],\Psi_{t}[\rho_2])\leq D(\rho_1,\rho_2)$.
Contractivity gives rise to the decrement of the trace distance under completely positive (CP) maps and is a key ingredient in the definition of non-Markovianity.

According to Ref.~\cite{BreuerPRL09}, a process is defined as non-Markovian if there is a pair of initial states $\rho_{1,2}(0)$ of the system and a time $t$ in their evolution such that 
\begin{equation}\label{trdistance}
\sigma(t,\rho_{1,2}(0))=\frac{d}{dt} D(\rho_1(t),\rho_2(t)){>}0.
\end{equation}
The associated measure of non-Markovianity is then calculated as $\mathcal{N}{=}\max_{\rho_{1,2}(0)}\int_{\sigma_+}dt\ \sigma(t,\rho_{1,2}(0))$, where $\sigma_+$ is the union of the temporal domains where $\sigma(t,\rho_{1,2}(0)){>}0$. The optimization over the pair of input states ensures that such a figure of merit faithfully reveals non-Markovianity.

\emph{Bound to non-Markovianity.-} We now provide the connection between the emergence of non-Markovianity as witnessed by Eq.~(\ref{trdistance}) and the evolution of SECs by formulating an upper bound to the derivative of the trace distance between two system states $\rho_{1}^S$ and $\rho_{2}^S$ at a generic instant of time $t$. Our result can be effectively formulated as the following

\noindent
\theo{ For any  quantum process described by a completely positive map with associated system-environment interaction ruled by the propagator $\hat U_{(t,0)}=e^{-i\hat Ht}$ we have
\begin{equation}\begin{split}\label{formulone}
\sigma\left({t,\rho_{1,2}^S}\right){\leq}&\frac{1}{2}\bigg(\min_{k=1,2}\left\Vert\mathrm{Tr}_E\left\{\left[\hat H,\rho_{k}^S(t)\otimes(\rho_1^E(t)-\rho_2^E(t))\right]\right\}\right\Vert\\
+&\left\Vert\mathrm{Tr}_E\left\{\left[\hat H,(\chi_1^{SE}(t)-\chi_2^{SE}(t))\right]\right\}\right\Vert\bigg)
\end{split}\end{equation}
where $\rho_{1,2}^S$ are arbitrary systems states, $\chi^{SE}_{j}(t)=\rho^{SE}_j(t)-\rho^S_j(t)\otimes\rho^E_j(t)$ are the SECs, and $\rho_{j}^{S(E)}(t){=}\mathrm{Tr}_{E(S)}[\hat U_{(t,t')}\rho_{j}^{SE}(t') \hat U^{\dagger}_{(t,t')}]$ (for any $t'{<}t$) is the system (environment) state at the time $t$. }

The technical steps needed to prove this statement are sketched in the Appendix. Here we focus on the significance and the implications of this Theorem, which is the central result of our study.
Eq.~(\ref{formulone}) bridges the occurrence of non-Markovianity (as revealed by a growing trace distance) with the dynamics of SECs and environmental distinguishability. The upper bound consists of two different contributions. 
The first term contains the information about the two different states of the environment. By decomposing the interaction Hamiltonian in terms of the eigen-operators $\hat A^S_{\alpha}$ and $\hat B^E_{\alpha}$, so that $\hat H{=}\sum_{\alpha} \hat A^S_{\alpha}{\otimes}\hat B^E_{\alpha}$, the first term can be recast into the form $||\ [\!\ \sum_\alpha \gamma_\alpha \hat A_\alpha,\rho_{2(1)}^S(t)]\ ||$ with $\gamma_\alpha{=}\pm\mathrm{Tr}_E\{B_\alpha (\rho_{1}^E(t){-}\rho_{2}^E(t))\}$. Here $\gamma$ clearly depends on the difference between the environment states.
The second term of the bound accounts for the presence and evolution of SECs. This term contains all the non-diagonal elements of the $S$-$E$ state, depends by definition on both the reduced state of the system and the total system-environment state, and is such that $\mathrm{Tr}_{S(E)}[\chi_{j}^{SE}(t')]=0$.

The bound in Eq.~(\ref{formulone}) shows that if SECs are not produced across the evolution and the environment in left in the same conditions regardless of the state of the system, the process is necessarily Markovian. This demonstrates the intimate connection between SECs and the changes in distinguishability of different input states in a non-Markovian evolution, broadening the view on the occurrence of such effects.
On the other hand, our upper bound states that the creation of correlations can be compatible with a Markovian dynamics.

An important point to stress is that the Theorem is formulated under the assumptions of zero {\it initial} SECs: the initial state of system and environment is factorized. A unification between witness of initial SECs and witness of non-Markovian dynamics is proposed and thoroughly discussed in Ref.~\cite{tocome}.

\emph{Physical Model.}---We now test Eq.~\eqref{formulone} against a physically motivated situation that will help us illustrate its deep physical implications. We consider the generation of SECs and their relation with non-Markovianity in a simple unidimensional quantum many-body system embodied by $N+1$ spin-$1/2$ particles (labelled $n=0,1,..,N$) mutually coupled via an XX model and subjected to a transverse magnetic field. We are interested in the bipartition consisting of the small system given by spin $0$ and the environment represented by the remaining $N$ particles. 
Assuming units such that $\hbar=1$ across the manuscript, the corresponding Hamiltonian model is $\hat H{=}\hat H_{SE}+\hat H_E$ with
\begin{equation}\begin{split}\label{XX}
& \hat H_{SE}=-2J_0(\hat\sigma_0^x \hat\sigma_{1}^x+ \hat\sigma_0^y \hat\sigma_{1}^y),\\
& \hat H_E=-2J\sum_{n=1}^{N-1}(\hat\sigma_n^x \hat\sigma_{n+1}^x+ \hat\sigma_n^y \hat\sigma_{n+1}^y)-2B\sum_{n=1}^{N} \hat\sigma_n^z,
\end{split}\end{equation}
where $\hat\sigma^{k}_n$ is the $k$-Pauli matrix ($k{=}x,y,z$) for particle $n$, $B$ is the amplitude of the magnetic field affecting $S$ and $J$ ($J_0$) is the inter-environment (system-environment) coupling strength. This model presupposes that the free evolutions of $S$ and $E$ are identical, thus allowing the passage to the interaction picture without the introduction of time-dependent coefficients. The open-system evolution of $S$ and its consequences for non-Markovianity have been analysed in Ref.~\cite{ApollaroPRA11}, where it was found that, for interaction times that are well within the recurrence time of the system (when any information propagating across the chain returns to $S$ after reaching the end of the chain), there is a working point defined by $(J_0/J,B/J)$ at which the measure of non-Markovianity ${\cal N}$ is null. As the optimization inherent in the definition of such measure is achieved for system states lying on the equatorial plane of the Bloch sphere~\cite{ApollaroPRA11}, we consider the input states $\rho^S_{1,2}(0)=\ket{\pm}\!\bra{\pm}$ with $|{\pm}\rangle{=}(\ket{0}\pm\ket{1})/\sqrt{2}$. The environment is initialized in $\rho^E_{1,2}(0)=\rho^E_{ini}=\otimes_{i=1}^{N-1}\ket{0}_i\!\bra{0}$, so that the total system-environment state is $\rho^{SE}_{1,2}(0)=\rho^S_{1,2}(0)\otimes\rho^E_{ini}$. 

\begin{figure}
\includegraphics[scale=0.44]{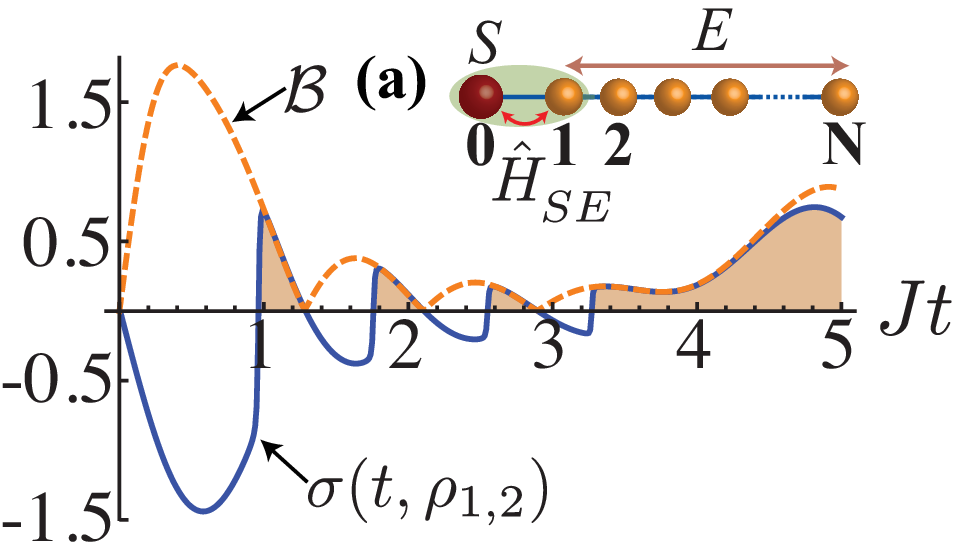}~~\includegraphics[scale=0.44]{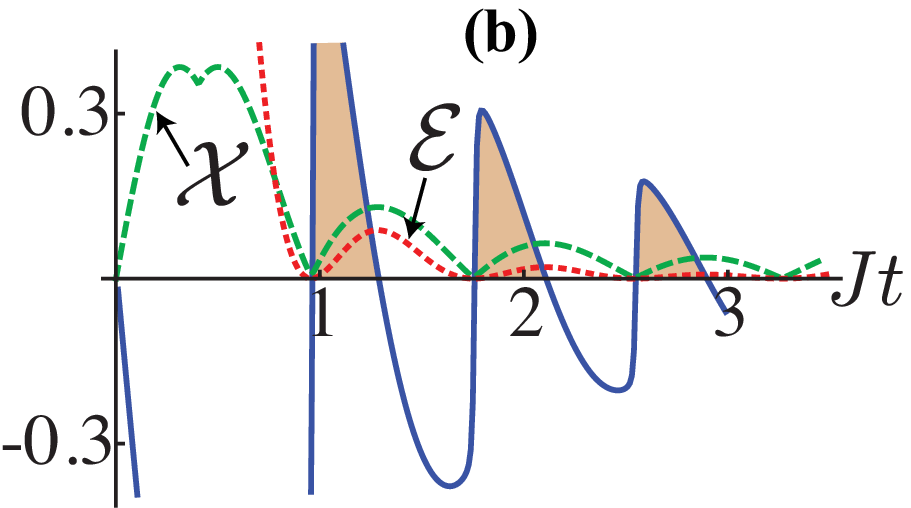} 
\caption{(Color online) {\bf (a)} We show $\sigma(t,\rho_{1,2})$ (solid blue line) and the right hand side of Eq.~\eqref{formulone} (which we label $\mathcal{B}$ for convenience and display as a dashed orange line) against the rescaled interaction time $Jt$ for a chain of $N=9$ particles ({\it i.e.} an environment with 8 spins) when $S$ is prepared in either $\ket{+}$ or $\ket{-}$ (see inset for a sketch of the physical situation at hand). {\bf (b)} We compare the behaviour of $\sigma(t,\rho_{12})$  (solid blue line) with the evolution of the environmental {\it indistinguishability} $\mathcal{E}{=}1-D(\rho^{E}_{1}(t),\rho^{E}_{2}(t))$ (amplified by a factor 10 for easiness of visualization, dotted red line) and the quantity $\mathcal{X}$ defined in the body of the manuscript (dashed green line). In both panels, $J_0/J{=}1$ and  $B=10^{-2}J$, while shaded regions highlight the regions of non-Markovianity.}
\label{LHS&RHS}
\end{figure}

Fig.~\ref{LHS&RHS} {\bf (a)} compares $\sigma(t,\rho_{1,2})$ and its upper bound under the above initial conditions and for $J_0/J{=}1,\ B/J=0.01$. We have restricted the width of the time-window that we consider to values within the time at which finite-size effects are expected to occur. This is easily estimated by considering that the maximum single-excitation group velocity of our model is $2$, which implies that it takes at least a time $N$ for a single excitation to leave $S$ and come back to it. As the right hand side of Eq.~(\ref{formulone}) is non negative, a quantitative comparison between the derivative of the trace distance and its upper bound is meaningful only when $\sigma(t,\rho_{1,2}(t))>0$. Interestingly, the bound becomes very tight as soon as the derivative of the trace distance becomes positive.

The two terms appearing in the bound have a different time behavior: the one depending on the reduced environmental states contributes the most to the right hand side. The second term, which is associated with the correlations, has local minima whenever  $\sigma(t,\rho_{1,2})$ crosses the horizontal axis. 
This observation prompts us to study the behavior of other quantities. In Fig.~\ref{LHS&RHS} {\bf (b)} we plot the evolution of the trace distance of the correlations ${\cal X}{=}D(\chi_{1}^{SE}(t),\chi_{2}^{SE}(t))$ and the {\it indistinguishability} between the environmental states ${\cal E}{=}1{-}D(\rho_{1}^{E}(t),\rho_{2}^{E}(t))$ 
finding that they are both minimal when $\sigma({t,\rho_{1,2}})$ changes sign from being negative, thus showing that non-Markovianity occurs when the environmental states become perfectly distinguishable [{\it i.e.} $D(\rho_{1}^{E}(t),\rho_{2}^{E}(t)){=}1$]. At these times, a careful analysis shows that not only ${\cal X}\simeq0$ but also the correlations within each state almost disappear ($||\chi_{1,2}^{SE}(t)||\simeq0$), thus leaving system-environment states that are basically factorized.

\begin{figure}
\includegraphics[scale=0.46]{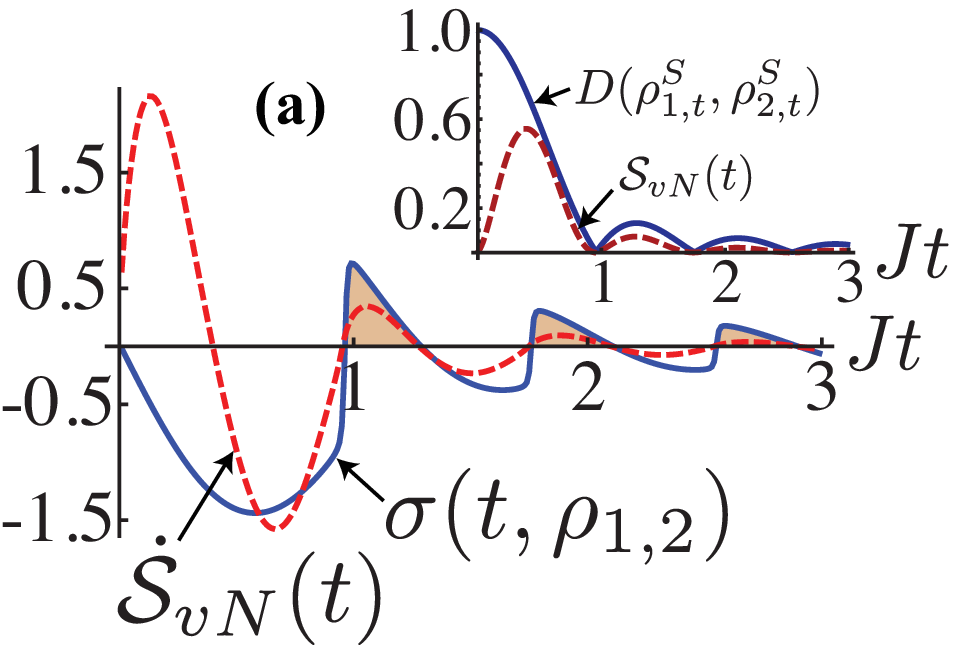}\includegraphics[scale=0.46]{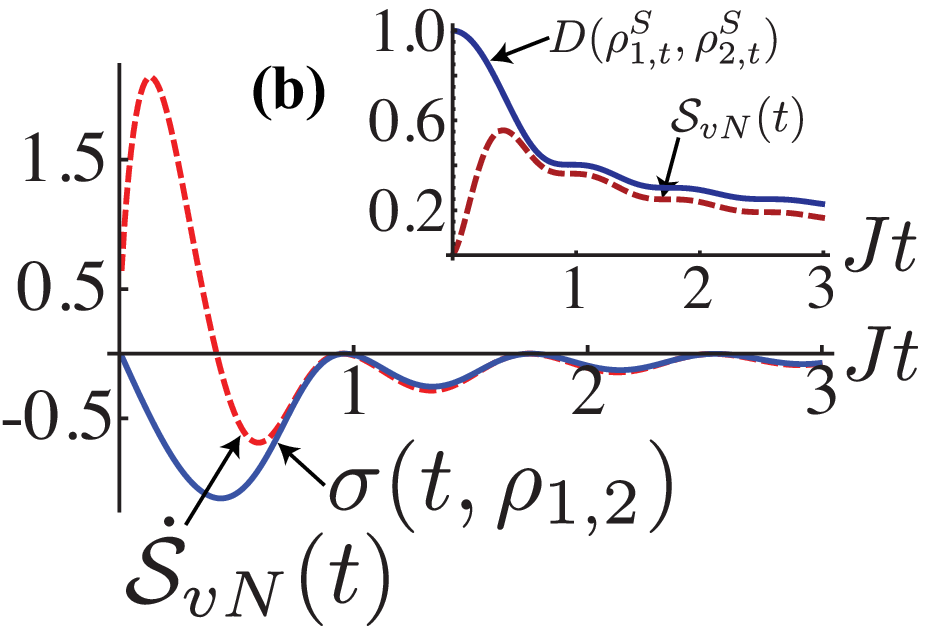}
\caption{(Color online) {\bf (a)} We show $\sigma({t,\rho_{1,2}})$ (solid blue line) and $\dot{\mathcal{S}}_{vN}(t)$ (dashed red curve) against the interaction time $Jt$ for $J_0/J=1,B/J=10^{-2}$. The trace distance $D(\rho_{1}^{S}(t),\rho_{2}^{S}(t))$ and entanglement associated to this situation are reported for the same time-window. The shaded regions highlight non-Markovianity. {\bf (b)} Same analysis but for the Markovian point corresponding to $J_0/J{=}1$ and $B/J{=}1/2$.}
\label{TrDist&EntSE}
\end{figure}

{\it Time-derivative of quantum mutual information.}--- We know pass to the quantification of SECs in non-Markovian dynamics. The quantum mutual information ($\mathcal{I}$) between $S$ and $E$ quantifies the total amount of information shared between system and environment, and accounts for classical and quantum correlations between the parts.
Given that the state of the total system (plus environment) is pure and evolves via unitary dynamics, the quantum mutual information is simply equal to twice the von Neumann entropy ${\cal S}_{vN}(t){=}{-}{\rm Tr}[\rho^S_{1,2}(t)\log_2\rho^S_{1,2}(t)]$ of the system (coinciding also with the entanglement between $S$ and $E$). Notice that the von Neumann entropy is the same for the two system's states. Fig.~\ref{TrDist&EntSE} {\bf (a)} compares the derivative of the von Neumann entropy with $\sigma({t,\rho_{1,2}})$: with the exception of the initial part of the dynamics, the two functions exhibit the same qualitative behaviour. When the evolution starts the quantum mutual information increases, so SECs are created, leading to a loss of information over the system's state and thus to decoherence. Later on, both mutual information and SECs decrease: this is the time-window where {\it re-coherence} takes place. If such a process becomes significant, and thus correlations become small enough (implying an actual comeback of the information that was previously encoded in the $S$-$E$ state to $S$ only), the system ``jumps" to the non-Markovian regime with an abrupt change of $\sigma({t,\rho_{1,2}})$. Quite interestingly, such re-coherence processes do not coincide with the non-Markovianity period, but seem to act as {\it precursors} of it.  

This behavior is typical of the working points at which the model at hand gives rise to non-Markovian processes certified by ${\cal N}{\neq}0$. On the other hand, as mentioned above, a parameter-regime exists for which the many-body dynamics corresponds to a perfectly Markovian process for spin $S$~\cite{ApollaroPRA11}. This is reported in Fig.~\ref{TrDist&EntSE} {\bf (b)}. Clearly, the much higher degree of quantum correlations between $S$ and $E$ at the times at which, in panel {\bf (a)}, the non-Markovian thresholds were passed, leaves ${\sigma}(t,\rho_{1,2}){<}0$. 
The quantum mutual information approaches zero (and $\sigma({t,\rho_{1,2}})$ is positive again) only for time-windows wide enough to cover the effects of to the chain's finite size. However, this sort of non-Markovianity should be set apart from the features discussed here and associated to shorter timescales. While the former are due to the physical return of excitations to the $S$ particle, the latter have a much deeper origin, now captured by Eq.~\eqref{formulone} and our analysis.

{\it Conclusive Remarks.}---We have proposed and analysed in depth a new approach to the establishment of non-Markovian dynamics, based on the assessment of the state of the environment and the explicit presence of SECs during the open evolution. Our perspective is that, in order to get insight into the reasons behind a non-Markovian process, one should focus on the structural modifications that the interaction with $S$ induces on the system-environment states. Quite significantly, our investigation appears to support such a vision by linking explicitly one of the measures for non-Markovianity to the presence of SECs and possible deviations of the environment from its initial state. We have provided an upper bound to the rate of change of the distinguishability between different states of $S$ that clearly shows a dependence on such key features. The tightness of the bound was investigated using a spin-chain system whose rich non-Markovianity diagram allowed the exploration of various interesting dynamical conditions~\cite{ApollaroPRA11} and the study of the subtle role played by SECs in the emergence of non-Markovianity. 

We believe that SECs can be instrumental also to address critically the interpretation of the measure of non-Markovianity based on the trace distance. 
Such a quantifier is operationally interpreted in terms of the so-called ``flow of information" to and from the environment: for non-Markovian systems, the natural direction of the decoherence process leading the system to lose energy/coherence/quantumness into the environment is, in certain time windows, reversed. The reversal of the flow of information is assumed to happen when the derivative of the trace distance increases.
Nevertheless, when used in the context of non-Markovian dynamics, the term information-flow does not have a uniquely defined mathematical meaning. 
Our view is that SECs embodies a proper figure of merit to define information flow. In such a perspective, information would be identified by the quantum mutual information $\mathcal{I}$ shared by system and environment, thus quantifying the total amount of SECs. On the other hand, its flow will be encompassed by the {time derivative} of $\mathcal{I}$. An {\it inflow} ({\it outflow}) of information to (from) the system would be signaled by  its negative (positive) time derivative. The application of such a definition to the physical model at hand would challange the commonly accredited view according to which $S$ recovers information while the trace distance is positive. In fact, in the physical model studied here, the re-coherence process, which is associated with an increased purity of the system's state and the decrease of SECs, occurs before the transition to a non-Markovian window and, in a sense, precedes it.

Our study calls out loud for a clearer definition of information flow in open quantum systems and proposes the us of correlations between system and environment as the figure of merit against which such an analysis should be performed.
L.M. and M.P. gratefully acknowledge A. Smirne and B. Vacchini for invaluable discussions. L.M. and C.R.R. thank the Centre for Quantum Technology for their kind hospitality during the early stages of this work. L.M. acknowledges support from EU under a Marie Curie IEF Fellowship (300368). M.P. thanks the UK EPSRC for support provided by a Career Acceleration Fellowship and a grant under the ÒNew Directions for EPSRC Research LeadersÓ initiative (EP/G004579/1). K.M. acknowledges the financial support by the National Research Foundation and the Ministry of Education of Singapore.

\section*{Proof of the stated Theorem}

We now provide the steps needed in order to derive the upper bound to $\sigma(t,\rho_{1,2})$ stated in Eq.~(\ref{formulone}). As discussed in the main body of the paper, we consider a quantum process described by a completely positive map and take the derivative of the trace distance between two system states, $\rho_{1}^S$ and $\rho_{2}^S$, at a generic instant of time $t$. We get 
\begin{equation}\label{ini}
\sigma({t,\rho_{1,2}})=\frac{1}{2}\lim_{t'\rightarrow t}\frac{||\rho_{1}^S(t)-\rho_{2}^S(t)||-||\rho_{1}^S(t')-\rho_{2}^S(t')||}{t-t'}.
\end{equation}
We now use the decomposition of the $S$-$E$ state at time $t'$ $\rho_{j}^{SE}(t')=\rho_{j}^{S}(t')\otimes\rho^{E}_{j}(t')+\chi_{j}^{SE}(t')$ and the triangular inequality for the trace distance to obtain
\begin{equation}\begin{split}\label{SecondStep}
||\rho_{1}^S(t)&-\rho_{2}^S(t)||\leq
||\mathrm{Tr}_E\{\hat U_{(t,t')}(\rho_{1}^S(t')-\rho_{2}^S(t'))\otimes\rho_{1(2)}^E(t')\hat U_{(t,t')}^{\dagger}\}||\\
&+\min_{k=1,2}||\mathrm{Tr}_E\{\hat U_{(t,t')}(\rho^S_{k}(t'))\otimes(\rho_{1}^E(t')-\rho_{2}^E(t'))\hat U_{(t,t')}^{\dagger}\}||\\
&+||\mathrm{Tr}_E\{\hat U_{(t,t')}(\chi_{1}^{SE}(t')-\chi_{2}^{SE}(t'))\hat U_{(t,t')}^{\dagger}\}||.
\end{split}\end{equation}
By using the fact that the evolution of a system through the unitary operator $\hat U_{(t,0)}$ is described by the completely positive (CP) dynamical map, the first term in the right hand side of Eq.~\eqref{SecondStep} is written as $||\tilde{\Phi}_{(t,t')}[\rho_{1}^S(t')]-\tilde{\Phi}_{(t,t')}[\rho_{2}^S(t')]||$, where $\tilde{\Phi}_{(t,t')}[\rho_{i}^S(t')]=\mathrm{Tr}_E\{U_{(t,t')}\rho_{i}^S(t')\otimes\rho_{1(2)}^E(t') U_{(t,t')}\}$ is a CP trace-preserving map. Using again contractivity we have that
\begin{equation}\label{ThirdStep}
||\tilde{\Phi}_{(t,t')}[\rho_{1}^S(t')]-\tilde{\Phi}_{(t,t')}[\rho_{2}^S(t')]||\leq||\rho_{1}^S(t')-\rho_{2}^S(t')||.
\end{equation}
By plugging Eq.~\eqref{ThirdStep} into Eq.~\eqref{SecondStep} and substituting back in Eq.~\eqref{ini}, we find
\begin{equation}
\begin{aligned}
\label{formulonerozzo}
&\sigma({t,\rho_{1,2}})\leq\frac{1}{2}\lim_{t'\rightarrow t}\bigg\{\frac{||{\Phi}_{(t,t')}[\rho_{1}^S(t')-\rho_{2}^S(t')]||-||\rho_{1}^S(t')-\rho_{2}^S(t')||}{t-t'}\\
&\left.+\frac{||\mathrm{Tr}_E\{\hat U_{(t,t')}(\chi_{1}^{SE}(t')-\chi_{2}^{SE}(t'))\hat U_{(t,t')}^{\dagger}\}||}{t-t'}\right.\\
&\left.+\min_{k=1,2}\frac{||\mathrm{Tr}_E\{\hat U_{(t,t')}\rho_{k}^S(t')\otimes(\rho_{1}^E(t')-\rho_{2}^E(t'))\hat U_{(t,t')}^{\dagger}\}||}{t-t'}\right\}.
\end{aligned}\end{equation}
The first term in the right hand side is non-positive due to the contraction property of the trace distance under CP maps. We discard it, thus providing a loser 
upper bound to $\sigma(t,\rho_{1,2})$. 
In order to manage the second and third terms we assume to know the Hamiltonian $\hat H$ regulating the $S$-$E$ interaction and expand $\hat U_{(t,t')}=e^{-i \hat H (t-t')}$ in power series, stopping it at first order in $\hat H$. We find
\begin{equation}
\begin{aligned}
\nonumber
&\lim_{t\rightarrow{t}'}\frac{||\mathrm{Tr}_E\{\hat U_{(t,t')}(\chi_{1}^{SE}(t'){-}\chi_{2}^{SE}(t'))\hat U_{(t,t')}^{\dagger}\}||}{(t-t')}\\
&=||\mathrm{Tr}_E \{[\hat H,\chi_{1}^{SE}(t){-}\chi_{2}^{SE}(t)]\}||.
\end{aligned}
\end{equation}

An analogous calculation holds for the third term, from which Eq.~(\ref{formulone}) is then straightforwardly found.

\end{document}